\documentclass{aa}     
\usepackage{amsmath}
\usepackage{amssymb}
\usepackage{graphicx}
\usepackage{natbib}
\usepackage{txfonts}
 
\begin{document}

%
     \title{HD 34700: A new T Tauri double-lined spectroscopic binary
\thanks{Based on observations obtained at the Haute-Provence Observatory,
  France.} \fnmsep \thanks{Table 2 is only available in electronic form at the CDS.}}
  \markboth{Abundances}{Arellano Ferro}
     \author{A. Arellano Ferro$^1$, Sunetra Giridhar$^2$}
     \institute{$^1$Instituto de Astronom\'{\i}a, Universidad Nacional Aut\'onoma de
  M\'exico, Apdo. Postal 70-264,
  M\'exico D.F. CP 04510; armando@astroscu.unam.mx \\
                $^2$Indian Institute of Astrophysics,  Bangalore 560034,
  India; giridhar@iiap.ernet.in\\}


 \abstract{We find the star HD 34700 to be a double-lined spectroscopic binary. 
  We also identify it as a  weak-line T Tauri object
  The spectra of both components are very similar and both show the Li I 
 feature at 6708 \AA. Strong arguments in favour of the binary nature of
 the star as opposed to other possibilities are offered. It is very
 likely that the companion is also a T Tauri star of similar mass.
 A projected rotational velocity, $v$sin$i$, of 25 and 23 km~s$^{-1}$ has been estimated for the blue and red components.
  We present a list of the lines identified and the radial velocities of the two
  components on three spectra obtained on consecutive nights.

  }

  %

  \maketitle

  \section{Introduction}
  The star HD 34700 (SAO 112630) has been known for its strong infrared excess since
  it was identified in the IRAS Point Source Catalogue (Odenwald 1986; Oudmaijer et
  al. 1992). Due to its infrared excess and its spectral classification as G0V, it
  became apparent that it was a pre-main
   sequence (PMS), probably a weak-line T Tauri star (wTTS). It has been included in
  numerous optical and infrared works on Vega-like stars. Most of these
  studies were made to analyse the observed Spectral Energy Distribution (SED) and
  derive the properties of the circumstellar material surrounding them (Sylvester et
  al. 1996; Sylvester and Skinner 1996; Eiroa et al. 2001).
  The star shows a $J-H$ colour compatible with that of a G0V star but excess in the
  $H-K$ colour, which is probably due to non-photospheric flux contributing to the $K$
  magnitude (Eiroa et al. 2001; Sylvester et al. 1996).
  Radio observations made by Zuckerman et al. (1995) have led to the detection
   of $CO$ for this object. These authors have reported the detection of $^{12}CO$
   as well as $^{13}CO$, V$_{\odot}$($CO$) of 21 km~s$^{-1}$, outer radius of molecular disk of 50~AU,
  and a dust mass of 1 Earth mass.
  The star also shows low level optical polarisation from which a non-spherical 
symmetry or disk-like geometry of the circumstellar dust is inferred (Bhatt \& Manoj 2000; Oudmaijer
  et al. 2001). Large grains (size $\sim$ 1 mm) are not abundant in the circumstellar material of
  HD 34700  but it is mostly made of small grains ($<$ 10 $\mu$m) (Sylvester and
  Skinner 1996). The star did not show variability in $JHK$ bands in the time scale of
  days or months (Eiroa et al. 2001).

   The first spectroscopic  analysis of this star was made by Mora et al. (2001).
  These authors have used seven  low resolution (6,700) spectra and
  two high resolution (49,000)  spectra to classify the star as a G0IVe.
 Other than emission lines they did not report any
  peculiarity in the spectrum of the star.

   Due to its position in the IRAS two-colour diagram (van der Veen \& Habing 1988)
 the star was included in our program on post-AGB candidates. However the general  
appearance of the spectrum is very  similar to those of some
 wTTS (moderate H${\alpha}$ emission and strong Li I 6708 \AA). 
While examining the spectra we found  a very  distinct
 line doubling of virtually all lines, including that of Li I line
 at 6708 \AA. In this paper, we report the duplicity of the star based on 
the analysis of the spectra taken on three consecutive nights.

  Much to our surprise the star is  continued to be listed as a post-AGB in the
  $SIMBAD$ data base.

  \section{Observations}

  Three spectra of HD 34700 were  obtained  during October 7-9, 2000 with the 1.93m
  telescope
  of the Haute-Provence Observatory (OHP), which is equipped with the high resolution
  (42,000) echelle
  spectrograph ELODIE. Details about the  performance and characteristics of the
  instrument have  been thoroughly described by Baranne et al. (1996).
   These spectra were reduced using spectroscopic
   data reduction tasks available in the IRAF package.

  \begin{figure}
  \includegraphics[width=8.cm,height=5.3cm]{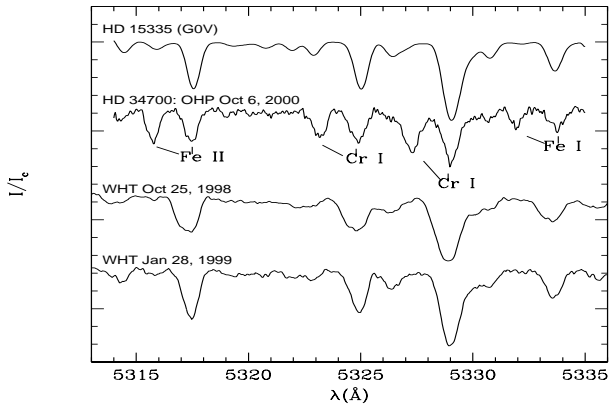}
  \caption{Comparison of HD 34700 OHP spectrum (second from the top) with that of
  13 Tri (HD 15335,  G0V) taken from Montes \& Martin (1998). The presence of
  line doubling is clearly seen. The two WHT EXPORT spectra at
the bottom do not show the line doubling but lines are broadened by the
 blending of the two components.}
  \end{figure}

  \begin{figure}
  \includegraphics[width=8.cm,height=5.3cm]{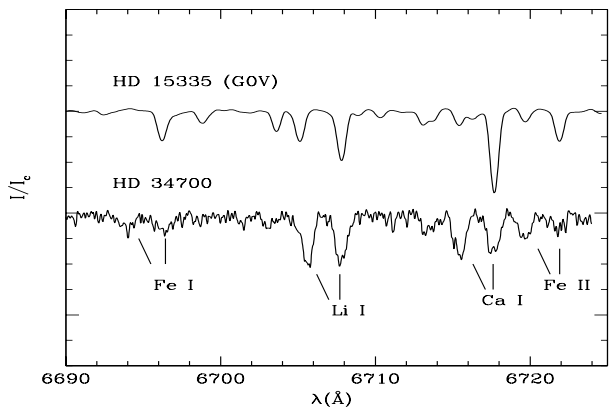}
  \caption{The Li 6708 \AA~ region in the OHP spectra compared with the standard star.}
\end{figure}

  \section{Spectral characteristics}

  We have found  line doubling in virtually all lines in the spectrum of HD 34700 from
  4000 to 6800 \AA ~in the three spectra taken on consecutive nights. 
The two components show clear radial velocity variations from night to night. 
We used the spectrum  of the highest S/N ratio (67 at 6000\AA) for line identification
  and line strength measurements. Two examples of the line doubling are shown in
  Figures 1 and 2, where we have compared the spectrum of HD 34700 with that of a
  well-known  G0V star  (HD 15335) from the high resolution spectral
   catalogue of Montes \& Martin (1998). We have degraded the resolution of
   HD 15335  such that it is comparable to the resolution of our spectrum. 
As the lines are double, they often blend with other
  stellar features, making the line identification difficult sometimes.
The identified lines and radial velocities of both
  the blue and the red components are listed in Table 2, available at the CDS.
We have
  included in Table 2 only those lines which
   have both the components clearly identified and unaffected by the neighbouring
   lines.
The table also contains the lower excitation potentials and the ratios of line strengths of the
   two components, which are consistently of similar strengths.
    Since line duplicity is shown irrespectively of
   where the lines are formed in the atmosphere,
  it is most likely that we have a composite spectrum of two different stars of
  similar spectral types.

   The mean blue and red radial velocities of the identified lines in the three spectra
are reported in Table 1. 

  In Figure 3 we display examples of double lines in the three spectra.  
Obvious radial velocity variations
  are seen on both line components which may be interpreted as due to the
  orbital motion of a binary system as discussed in section 4.
The line depths are affected by the veiling phenomenon known to occur
   in  T Tauri stars, consisting of additional flux coming from a
   featureless continuum that fills in photospheric absorption lines
   (Hessman \& Guenther 1997; Mora et al. 2001). 
  Thus an attempt to estimate the effective temperature and gravity of the star,
   by the standard procedure of excitation and ionisation equilibrium
   for  a set of Fe I and Fe II lines  with the same radial velocity,
    proved to be futile. It also prevented us from deriving the rotation 
  velocities by comparing synthetic line profiles  with observed ones.

 Very distinct variations in H${\alpha}$ profiles are observed in the three spectra in
 our possession and are
displayed in Fig. 4. While H${\alpha}$ variations  could be caused by changes
 in the  stellar chromospheric activity, it is also possible that veiling
 variations due to temperature changes do contribute to the observed changes
 in  H${\alpha}$. The weakening of the lines  due to veiling has  
 been illustrated by Mora et al. (2001) for five PMS stars.    

    All other lines also show depth variations from night to night, as it is illustrated
 for a few lines  in Figure 3. Therefore variations in the time scale of hours,
  like those described for the highly veiled T Tauri stars DR Tau, DG Tau and DI Cep
    (Hessman \& Guenther 1997) may also be present in HD 34700.

  \begin{figure}
  \includegraphics[width=8.cm,height=11.cm]{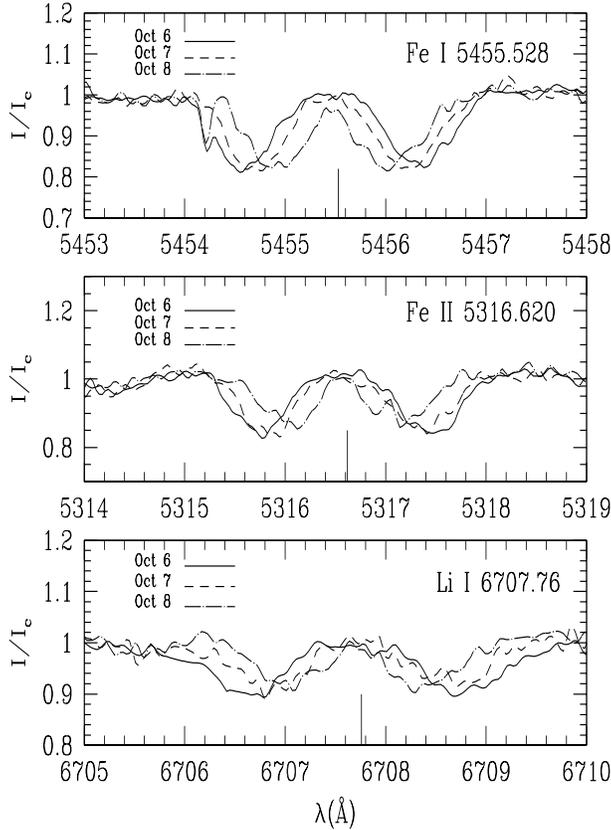}
  \caption{Three examples of line radial velocity variations due
  to the binary's orbital motion. The long vertical mark
 indicates the rest position of the line.
 Note the line cores getting closer with increasing date.}
  \end{figure}

  \section{Discussion}
  \label{sec:Discussion}
   Apart from the variations in the line strengths as described in the last
   section,  radial velocity variations are also evident.
   We have reasons to believe that the observed line doubling is caused by
   the presence of a companion around HD 34700.

  \subsection{The double-lined spectroscopic binary interpretation}

   Line doubling can be caused by the passage of a shock, like those
  observed in pulsating stars.  However, in that case
   the line splitting is not as large as it is  observed in Fig. 3
 and it is generally accompanied by large luminosity variations (Tsesevich 1975).

  We have also debated with the alternative explanation in terms of central
  emission in lines, giving the illusion of line doubling. But in that 
   case,  at the epoch of weaker emission, one would expect the
    two absorption components to appear stronger and closer. The observed
   line components do not follow this pattern but the line strengths
   of component pairs remained unchanged over the three epochs, although
   their separation changes. Secondly, if the two components are treated as
   part of a single wide absorption separated by central emission, then the
   FWHM of this single wide absorption turns out to be larger than 2 \AA.
   If this broadening is ascribed to rotation, it would require $v$sin$i$ larger than
   100 km~s$^{-1}$ but for early G type PMS, the observed rotational velocities
do not exceed about 40 km~s$^{-1}$ (Mora et al. 2001).
  Furthermore, as neither a P-Cygni profile is observed in H${\alpha}$ nor other emission lines
   are present, the line doubling is not likely to be caused by a central
   emission. 

  \begin{table}[htbp]
  \footnotesize{
  \caption{Heliocentric radial velocities in HD 34700 and its companion}
  \begin{center}
  \begin{tabular}{ccccc}
  \noalign{\smallskip}
  \hline
  \noalign{\smallskip}
  \noalign{\smallskip}
  \multicolumn{1}{c}{Date}&
  \multicolumn{1}{c}{$V_{r,\odot}$ (blue)}&
  \multicolumn{1}{c}{$V_{r,\odot}$ (red)}&
  \multicolumn{1}{c}{N}&
  \multicolumn{1}{c}{HJD}\\
  \noalign{\smallskip}
  \multicolumn{1}{c}{}&
  \multicolumn{1}{c}{km~s$^{-1}$}&
  \multicolumn{1}{c}{km~s$^{-1}$}&
  \multicolumn{1}{c}{}&
  \multicolumn{1}{c}{(245 1000.+)}\\

  \noalign{\smallskip}
  \hline
  \noalign{\smallskip}
  Oct 6, 2000 & -27.0$\pm$ 4.3  & +69.3$\pm$ 3.9 & 106 & 823.70130\\
  Oct 7, 2000 & -21.0$\pm$ 5.8  & +63.6$\pm$ 5.8 &  98 & 824.72364\\
  Oct 8, 2000 & -9.5$\pm$ 4.6  & +53.4$\pm$ 5.8 &  80 & 825.66518\\
              \noalign{\smallskip}
              \hline
              \noalign{\smallskip}
  \end{tabular}
  \end{center}
  }
N= number of lines included in the calculation.
  \end{table}

    We are therefore more inclined to believe that the two components
    are arising from two different stars. Considering that the line strengths are
    very similar and that the Li I 6708 \AA ~feature appears double and also follows 
the velocity pattern exhibited by other photospheric  lines,  it is likely that
    both members of the binary system are T Tauri stars of similar masses.
    This phenomenon of line doubling in HD 34700 has somehow escaped attention in previous works.

  \begin{figure}[tbp]
  \includegraphics[width=8.cm,height=4.cm]{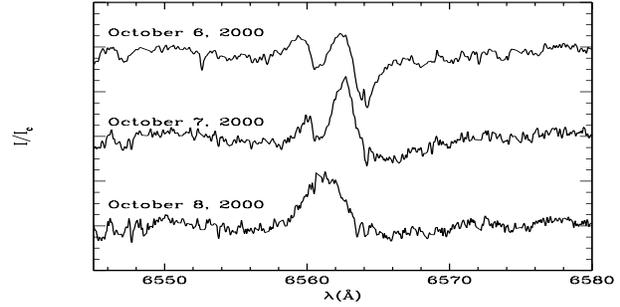}
  \caption{ H$\alpha$ line variations in three consecutive nights. It can be seen how
  the line is being filled in. The equivalent width of H$\alpha$ in the bottom spectrum is 0.6 \AA.}
  \end{figure}

 We  have inspected the high resolution
spectra, obtained on October 25, and January 28, 1999 using the 4.2m 
William Herschell Telescope (WHT) at La Palma as part of the EXPORT project (e.g. Mora et al. 2001). 
As an example we have included in Fig. 1. a portion of the two WHT spectra.
Line doubling cannot be observed but only a line broadening, which is different
for the two epochs. Those spectra seem to have been obtained at orbital phases
when the lines from the two components were almost merged, hence line
 duplicity was missed even at the high resolution of the WHT spectra. 
Unfortunately these two spectra do not cover the H$\alpha$ and 
Li I 6708 \AA ~regions.

  A list of the known PMS spectroscopic binaries includes forty systems (Melo et al. 2001). The orbital periods are between 1.6 and 2500 days, but 80\% have
  periods smaller than 100 days and about 50\% are double-lined. They are mostly of 
K spectral type but F, G and K  combinations also occur. HD 34700 should be
  added to the list. With spectral type of G0,
the star is among the hottest known wTTS double-lined spectroscopic binaries.

  \subsection{Rotational velocity}
 Due to the reasons mentioned in section 3, we did not use the 
 method of comparing synthetic and observed line profiles
to estimate the rotation velocities for the two components. Instead, we
 followed the FWHM vs. total broadening calibration described in Fekel (1997).
 We found the mean FWHM of 0.80 and 0.75 \AA ~for the blue and red components.
 The FWHM of the instrumental profile (derived from weak telluric lines)
 was 0.15 \AA ~corresponding to broadening of 8.5 km~s$^{-1}$, and for spectral
 type GO IV, a macroturbulence velocity of 4.0 km~s$^{-1}$ was adopted.
 After subtracting the contribution from the above mentioned broadenings we
 have derived rotation velocities of 25 and 23 km~s$^{-1}$ for the
 blue and red components  present in the spectrum.
 Following the discussion on the errors given in  Fekel (1997), these estimates
 may have an uncertainty of $\pm$ 1 km~s$^{-1}$.
 These values are considerably smaller than  46 km~s$^{-1}$
  determined by Mora et al. (2001) but their result has clearly been 
affected by line blending. With this rotational velocities,
 and assuming a prominent chromospheric activity, light variations
 with periodicities of the order of a few days are
expected. A photometric campaign on the star would be most valuable
 to test this scenario.

  \subsection{Asymmetric dust envelope}

  The $K$ image of the star in the 2MASS catalogue is shown in Figure 5.
  The asymmetry of the emission is evident,  similar to that in $J$ and $H$ bands.
  The size of the IR emitting
  region in the NE-SW direction is about 20 seconds of arc. Adopting the distance
  to the star 180 pc as a lower limit (van den Ancker et al. 1998), the minimum
  physical size is about 30 000 AU and about half that size in the orthogonal
  direction.
 At these scale the non-symmetrical distribution of the circumstellar dust, is most 
probably due to the
  existence of more than one star near the core, although to be conclusive on 
this topic will have 
to wait until the orbit of this binary star is duly parametrized.

  \begin{figure}
  \begin{center}

  \includegraphics[width=6.cm,height=6.cm]{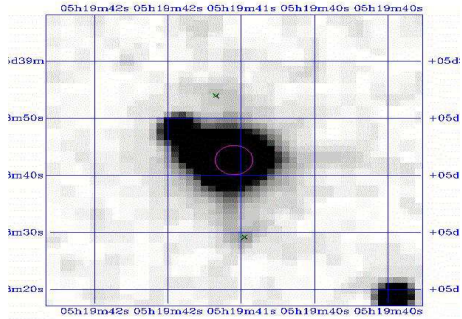}
  \caption{$K$ image of HD 34700 from the 2MASS catalogue. North is up and East is left. The asymmetry of the
  IR emitting region is likely due to the presence of a double star in the core
  of the region.}
  \end{center}

  \end{figure}

  \section{Conclusions}

  HD 34700 is a double-lined  spectroscopic binary. Since it shows prominent Li I, moderate H$\alpha$ emission and strong IR emission the star can safely be
  classified as a weak-line T Tauri star.

  Because of the similarity of the primary and secondary spectra and the
   presence of Li I 6708 \AA ~in both components, we believe that
  this double-lined spectroscopic binary is harbouring two T Tauri stars of similar masses.
A spectrum very similar to that of HD 34700 is displayed by Covino et al. (2001) 
for the K2+K2 PMS system RX J0530.7-0434.

  There are 26 known double-lined PMS spectroscopic binaries (Melo et al. 2001). HD 34700 should 
be added to the growing list of double-lined T Tauri spectroscopic binares.

  With only few spectra to our disposal we cannot go further on
   orbital analysis, but the  continuous spectral monitoring at  high resolution
    is urgently required to confirm our finding and to derive the orbital parameters
   for this object.

  \acknowledgements
  We are thankful to Drs. Carlos Eiroa and Benjamin Montesinos for making the two EXPORT WHT high resolution spectra available to us. AAF acknowledges support from DGAPA-UNAM grant through project IN110102 and is
  thankful to the CONACyT (Mexico) and the Department of Science and Technology
  (India), for the travel support and local hospitality respectively under
  Indo-Mexican collaborative project
  DST/INT/MEXICO/RP001/2001.

  This publication makes use of data products from the Two Micron All Sky Survey,
  which is a joint project of the University of Massachusetts
  and the Infrared Processing and Analysis Center/California Institute of Technology,
  funded by the National Aeronautics and Space
  Administration and the National Science Foundation
  %

  \end{document}